\documentstyle[aps,twocolumn]{revtex}
\begin{document}
\title{Chern-Simons-fermion model of quarks}
\author{Ernesto A. Matute\thanks{E-mail address: ematute@lauca.usach.cl}}
\address{Departamento de F\'{\i}sica, Universidad de Santiago de
Chile, Casilla 307, Santiago 2, Chile}
\maketitle
\begin{abstract}
We propose an extension of the standard model where quarks are viewed as
fermions with a ``bare'' integer (weak) hypercharge which is normalized
with a fractional part created by a quantized topological
Chern-Simons configuration of the weak gauge fields.  Consistency with
hypercharge patterns not included in the standard model is shown.
\end{abstract}

It is well known that the standard model of the strong, weak and
electromagnetic interactions \cite{Wein}, in spite of its spectacular
success, cannot be complete.  The recent experimental indications
for nonzero neutrinos masses \cite{Groom} are manifestations of physics
beyond the standard model.  It means that right-handed components of
neutrinos have to be introduced which are completely neutral under
the standard gauge group
$\mbox{SU(3)}_{c}\mbox{xSU(2)}_{L}\mbox{xU(1)}_{Y}$ \cite{Branco}.
Although they can be inserted anyway as sterile components if neutrinos
are massless.  Hence, a pattern of left- and right-handed quarks and
leptons is exhibited with (weak) hypercharges in their first generation
related as
\begin{eqnarray}
& & Y(u_{L}) = Y(\nu_{L}) + \frac{4}{3} =
- 1 + \frac{4}{3} ,  \nonumber \\
& & \nonumber \\
& & Y(u_{R}) = Y(\nu_{R}) + \frac{4}{3} = \;
0 + \frac{4}{3} ,  \nonumber \\
& & \nonumber \\
& & Y(d_{L}) = Y(e_{L}) + \frac{4}{3} =
- 1 + \frac{4}{3} ,  \nonumber \\
& & \nonumber \\
& & Y(d_{R}) = Y(e_{R}) + \frac{4}{3} =
- 2 + \frac{4}{3} ,        \label{hyper}
\end{eqnarray}
and electric charges connected by
\begin{eqnarray}
& & Q(u) = Q(\nu) + \frac{2}{3} = \; 0 + \frac{2}{3} ,  \nonumber  \\
& & \nonumber \\
& & Q(d) = Q(e) + \frac{2}{3} = - 1 + \frac{2}{3} , \label{elect}
\end{eqnarray}
where the conventional normalization $Q=T_{3}+Y/2$ has been adopted. The
second and third generation follow exactly the same scheme.  If these
equations are considered as prescriptions for the charge structures of
confined quarks, instead of leptons, the quark hypercharge, as shown in
Eq. (\ref{hyper}), can be viewed as the sum of an integer part, which
is equal to that of its lepton ``partner'', and a fractional part which
appears as a global hypercharge independent of flavor and handness, and
conserved by strong and electroweak interactions.  Moreover, the equality
between the integer part of quark hypercharges and lepton hypercharges
may reflect a discrete electroweak ${\cal Z}_{2}$ symmetry between the
quark and lepton families, but at a subquark or ``bare'' level. Such
patterns are not explained or included in the standard model, so that an
extended model is demanded.

Preon models of quarks and leptons have been proposed at the level of
accounting for the pattern of Eq. (\ref{elect}) \cite{Harari}.  However,
to understand Eq. (\ref{hyper}), it would mean to assume that the
electroweak symmetry below the composite scale is
$\mbox{SU(2)}_{L}\mbox{xU(1)}_{Y}$, with fundamental electroweak gauge
bosons and a new superstrong binding force implying proliferation of
preons, exotic fermions, and composite particles.

The purpose of this Letter is to show that the above features exhibited
by quarks may be understood alternatively in terms of
Chern-Simons-fermions, in the sense that a quark may be pictured as a
``bare'' fermion with integer hypercharge, a sort of predecessor or
pre-state of the quark that we will refer to as the prequark
\cite{Matute}, ``dressed'' or normalized with a fractional hypercharge
created by a topological Chern-Simons configuration of the weak gauge
fields \cite{Bertl}, which is introduced to cancel the gauge anomalies
produced by the prequark current.  In this model, the prequark has the
spin, color charge, and flavor of the quark, without existence of new
strong interactions as in preon models of quarks and leptons.  Leptons
are assumed to be absolutely elementary.  Thus, aside from the quark
hypercharge split, all the aspects of the standard model are retained.

To introduce the Chern-Simons-fermions properly, we follow a series of
steps which combines prequarks, leptons,
$\mbox{SU(3)}_{c}\mbox{xSU(2)}_{L}\mbox{xU(1)}_{Y}$ gauge theory, and
Chern-Simons field theory.  We start by postulating that the primary
fermionic constituents of matter are prequarks, which here will be
denoted by hats, and leptons.  Table 1 gives a classification of
their first generation according to the representations they furnish of
the standard gauge group, in which right-handed fermions have been
replaced by their left-handed charge-conjugate partners.
A ${\cal Z}_{2}$-symmetry between prequarks and leptons in the
electroweak sector is explicitly exhibited.

The assignment of integer hypercharges for prequarks, however, introduces
triangle gauge anomalies of the $\mbox{U(1)x[SU(2)]}^{2}$ and
$\mbox{[U(1)]}^{3}$ type.  Instead of adding exotic fermions to cancel
these anomalies, as it is usually done \cite{Georgi}, we propose an
approach with anomaly compensation terms involving topological
Chern-Simons configurations of gauge fields.  These local counterterms
would restore gauge symmetry and current conservation via quark
formalism.  The prequark hypercharges given in Table 1 are then ``bare''
values.  What can we gain with this approach?  Firstly, an unpleasant
proliferation of basic fermions is avoided.  Secondly, these anomaly
cancellation counterterms may bring about the required topological
fractional hypercharges to construct quarks from prequarks.  And thirdly,
a fundamental discrete symmetry between prequarks and leptons may be
revealed in the electroweak sector.  We will show that this is actually
the case and that the standard model is then consistently obtained.

Following the standard work \cite{Wein}, we find that the U(1) gauge
current
\begin{equation}
\hat{J}^{\mu}_{Y} = \overline{\hat{f}}_{qL} \gamma^{\mu} \frac{Y}{2}
\hat{f}_{qL} + \overline{f}_{\ell L} \gamma^{\mu} \frac{Y}{2}
f_{\ell L} ,
\end{equation}
with $\hat{f}_{qL}$ and $f_{\ell L}$ uniting all the left-handed
prequarks and leptons of one generation, respectively, exhibits the
$\mbox{U(1)x[SU(2)]}^{2}$ and $\mbox{[U(1)]}^{3}$ anomalies
\begin{eqnarray}
\partial_{\mu} \hat{J}^{\mu}_{Y} &=& - \frac{g^{2}}{32 \pi^{2}}
\biggl( \sum_{\hat{q}_{L} , \ell_{L}} \frac{Y}{2} \biggr)
tr W_{\mu\nu} \tilde{W}^{\mu\nu}   \nonumber \\
& & \nonumber \\
& & - \frac{{g'}^{2}}{48 \pi^{2}}
\biggl( \sum_{\hat{f}_{L}} \biggl(\frac{Y}{2}\biggr)^{3} \biggr)
F_{\mu\nu} \tilde{F}^{\mu\nu} ,      \label{div1}
\end{eqnarray}
where
\begin{eqnarray}
& & W_{\mu\nu} = \tau^{a} W^{a}_{\mu\nu} , \nonumber \\
& & \nonumber \\
& & \tilde{W}_{\mu\nu} = \frac{1}{2} \epsilon_{\mu\nu\lambda\rho}
\tau^{a} W^{a \lambda\rho} , \nonumber \\
& & \nonumber \\
& & W^{a}_{\mu\nu} = \partial_{\mu} W^{a}_{\nu} -
\partial_{\nu} W^{a}_{\mu} + g \epsilon^{abc} W^{b}_{\mu} W^{c}_{\nu} ,
\nonumber \\ & & \nonumber \\
& & F_{\mu\nu} = \partial_{\mu} A_{\nu} - \partial_{\nu} A_{\mu} ,
\;\;\;\; \tilde{F}_{\mu\nu} = \frac{1}{2} \epsilon_{\mu\nu\lambda\rho}
F^{\lambda\rho} ,
\end{eqnarray}
and $g$ and $g'$ are the SU(2) and U(1) gauge coupling constants,
respectively.  The first sum of Eq. (\ref{div1}) runs over
\linebreak
%-------------------------------------------------------------------
\begin{table}
  \caption{First generation of prequarks and leptons classified
   according to the $\mbox{SU(3)}_{c}\mbox{xSU(2)}_{L}\mbox{xU(1)}_{Y}$
   representation.}
  \begin{tabular}{lccr}
  Fermions & $\mbox{SU(3)}_{c}$ & $\mbox{SU(2)}_{L}$ & $\mbox{U(1)}_{Y}$
  \\  \hline
  $\hat{q}_{L} = \left( \begin{array}{cc}
                       \hat{u}_{L} \\ \hat{d}_{L}
                       \end{array}
                \right)$  & 3 & 2 & -1 \\
  $\hat{u}^{c}_{L}$ & $\bar{3}$ & 1 & 0 \\
  $\hat{d}^{c}_{L}$ & $\bar{3}$ & 1 & 2 \\
  $\ell_{L} = \left( \begin{array}{cc}
                       \nu_{L} \\ e_{L}
                       \end{array}
                \right)$  & 1 & 2 & -1 \\
  $\nu^{c}_{L}$ & 1 & 1 & 0 \\
  $e^{c}_{L}$ & 1 & 1 &  2
  \end{tabular}
\end{table}
%-------------------------------------------------------------------
\noindent
the fermions in the doublets, with the hypercharge $Y$ in units of $g'$;
the second one does over all prequarks and leptons.  The anomalies are
introduced because these sums do not vanish in each generation:
\begin{equation}
\sum_{\hat{q}_{L}, \ell_{L}} Y = - 8 , \;\;\;\;
\sum_{\hat{f}_{L}} Y^{3} = 24 .
\label{sums}
\end{equation}

As frequently noted in standard references \cite{Wein}, the terms
on the right-handed side of Eq. (\ref{div1}) are divergences of
gauge-dependent currents:
\begin{eqnarray}
\partial_{\mu} \hat{J}^{\mu}_{Y} &=& -
\biggl( \sum_{\hat{q}_{L} , \ell_{L}} \frac{Y}{2} \biggr)
\frac{\partial_{\mu} K^{\mu}}{2}  \nonumber \\
& & \nonumber \\
& & - \biggl( \sum_{\hat{f}_{L}} \biggl(\frac{Y}{2}\biggr)^{3} \biggr)
\frac{\partial_{\mu} L^{\mu}}{2}        \label{div2}
\end{eqnarray}
with
\begin{eqnarray}
K^{\mu} &=& \frac{g^{2}}{8 \pi^{2}} \epsilon^{\mu\nu\lambda\rho}
tr (W_{\nu} \partial_{\lambda} W_{\rho} - \frac{2}{3} i g W_{\nu}
W_{\lambda} W{\rho}) , \nonumber \\
& & \nonumber \\
L^{\mu} &=& \frac{{g'}^{2}}{12 \pi^{2}} \epsilon^{\mu\nu\lambda\rho}
A_{\nu} \partial_{\lambda} A_{\rho} .    \label{CS}
\end{eqnarray}
These quantities, often referred to as Chern-Simons classes or
topological currents, are not invariant under SU(2) and U(1) gauge
transformations, respectively.  Nevertheless, they can be combined with
the anomalous fermionic current $\hat{J}^{\mu}_{Y}$ to define a new
current
\begin{equation}
J^{\mu}_{Y} = \hat{J}^{\mu}_{Y} +
\biggl( \sum_{\hat{q}_{L} , \ell_{L}} \frac{Y}{2} \biggr)
\frac{K^{\mu}}{2} +
\biggl( \sum_{\hat{f}_{L}}
\biggl(\frac{Y}{2}\biggr)^{3} \biggr) \frac{L^{\mu}}{2} ,
\label{curr1}
\end{equation}
which is conserved and free of anomalies.  We remark that the bosonic
fields are not beset by anomalies, but serve to remove the anomalies from
fermionic currents.  Also, the local counterterms to be added to the
Lagrangian of prequarks and leptons, needed to obtain the anomaly-free
current of Eq. (\ref{curr1}), are
\begin{equation}
\Delta {\cal L} = \frac{g'}{2}
\biggl( \sum_{\hat{q}_{L} , \ell_{L}} \frac{Y}{2} \biggr)
K^{\mu} A_{\mu} ,      \label{Lagran}
\end{equation}
where only the non-Abelian Chern-Simons current of Eq. (\ref{CS}) is
required; the counterterms with the Abelian current vanish because of
the antisymmetry of $\epsilon^{\mu\nu\lambda\rho}$, so that one can
always introduce this current in Eqs. (\ref{Lagran}) and (\ref{curr1}).

The charge corresponding to the current of Eq. (\ref{curr1}) is
\begin{eqnarray}
Q_{Y} = \int d^{3}x J^{o}_{Y} &=& \int d^{3}x \biggl[ \hat{J}^{o}_{Y} +
\biggl( \sum_{\hat{q}_{L} , \ell_{L}} \frac{Y}{2} \biggr)
\frac{K^{o}}{2}  \nonumber \\ & & \nonumber \\ & &
+ \biggl( \sum_{\hat{f}_{L}} \biggl(\frac{Y}{2}\biggr)^{3} \biggr)
\frac{L^{o}}{2} \biggr] .
\label{charge}
\end{eqnarray}
This charge is not gauge invariant because of the existence of the
topological charge associated with the non-Abelian gauge fields;
in the case of the Abelian fields, a gauge transformation causes a
surface integral that vanishes at infinity.  Moreover, it is not
conserved. This can be seen by integrating $\partial_{\mu} J^{\mu}_{Y}$
over Euclidean 4-space to obtain a change that can be presented in the
form \cite{Huang1}
\begin{equation}
Q_{Y}(\infty) = Q_{Y}(-\infty) -
\biggl( \sum_{\hat{q}_{L} , \ell_{L}} \frac{Y}{2} \biggr)
\frac{Q_{K}}{2} ,                      \label{bounds}
\end{equation}
where
\begin{equation}
Q_{K} =  \int \partial_{\mu}K^{\mu} d^{4}x
= \frac{g^{2}}{16 \pi^{2}} \int
tr (W_{\mu\nu} \tilde{W}^{\mu\nu}) d^{4}x ,     \label{char1}
\end{equation}
is the so-called topological charge which can have nonzero values in
Euclidean space. In fact, it holds that \cite{Huang2}
\begin{equation}
Q_{K} = n ,  \label{nchar}
\end{equation}
where the integer number $n$ is named the Pontryagin index or winding
number and it is defined by the global characteristics of the gauge
field.  Thus, the current of Eq. (\ref{curr1}) is really not
conserved.  However, quark formation solves this serious problem as
we now show.

The anomaly-free current $J_{Y}$ of Eq. (\ref{curr1}) can be separated
into the prequark and lepton parts, as follows:
\begin{eqnarray}
J^{\mu}_{Y_{q}} &=& \overline{\hat{f}}_{qL} \gamma^{\mu}
\frac{\hat{Y}_{q}}{2} \hat{f}_{qL} +
\biggl( \sum_{\hat{q}_{L} , \ell_{L}} \frac{Y}{2} \biggr)
\frac{K^{\mu}}{2} + \biggl( \sum_{\hat{f}_{L}}
\biggl(\frac{Y}{2}\biggr)^{3} \biggr) \frac{L^{\mu}}{2} ,
\nonumber \\ & & \nonumber \\
J^{\mu}_{Y_{\ell}} &=& \overline{f}_{\ell L} \gamma^{\mu}
\frac{Y_{\ell}}{2} f_{\ell L} ,
\label{curr12}
\end{eqnarray}
so that the prequark current concentrates all the gauge field
contributions and the lepton one retains its standard form, as assumed
from the beginning.  We note that the modified prequark current is a
collective quantity that involves prequarks, leptons, and gauge fields.
The change of total prequark hypercharge is obtained from Eqs.
(\ref{bounds}) and (\ref{nchar})
\begin{equation}
Y_{q}(\infty) = Y_{q}(-\infty) - \biggl(
\sum_{\hat{q}_{L} , \ell_{L}} Y \biggr) \frac{n}{2} .
\label{char2}
\end{equation}
We next define a final hypercharge for {\it each} left-handed prequark
according to
\begin{equation}
Y_{q} = \hat{Y}_{q} - \biggl( \sum_{\hat{q}_{L} , \ell_{L}} Y \biggr)
\frac{n}{2\hat{N}_{q}} = \hat{Y}_{q} + \frac{n}{3} ,
\label{char12}
\end{equation}
where we have divided the nonzero value of the whole hypercharge
variation by the total number of left-handed prequarks,
$\hat{N}_{q}=12$, in each generation.  Here, since there is no other
restriction to impose, we have advocated the principle of equality or
democracy at the local level for all the prequarks of the system.
In a sense, it appears that the original integer-charged prequarks
effectively ``swallow'' equal fractions of the global topological charge,
in this manner transforming themselves into final prequarks with a
greater fractional charge.

Now, working backwards, the extra hypercharge $\Delta Y = n / 3$
should be counted as itself a part of the original prequark hypercharge.
To make the calculation self-consistent, the assumed hypercharge should
already contain this topological piece.  In other words, the final values
obtained from Eq. (\ref{char12}) are the actual hypercharges that we
should apply.  Going back to the above Eqs. (\ref{div1}), (\ref{div2}),
(\ref{curr1}), (\ref{Lagran}), (\ref{charge}), (\ref{bounds}),
(\ref{curr12}), (\ref{char2}), and (\ref{char12}), we see that the
final hypercharges automatically cancel the gauge anomalies of the
current and makes it to be gauge invariant and conserved if the
Pontryagin index for the non-Abelian gauge fields turns to be
\begin{equation}
n = 4 .
\label{n}
\end{equation}
In fact, changing the original ``bare'' hypercharge $\hat{Y}_{q}$ of
prequarks at the distant past by the final physical ones $Y_{q}$ at the
distant future, we have
\begin{equation}
\sum_{q_{L}, \ell_{L}} Y = 0 , \;\;\;\;
\sum_{f_{L}} Y^{3} = 0 ,
\end{equation}
instead of the values given in Eq. (\ref{sums}), and the physical
current is then
\begin{equation}
J^{\mu}_{Y_{q}} = \overline{f}_{qL} \gamma^{\mu} \frac{Y_{q}}{2} f_{qL} ,
\label{equiv}
\end{equation}
where $f_{q}$ describes final prequarks.  We should note that actually
all anomalies cancel within each generation of prequarks and leptons
for the above value of $n$.  Explicit solutions for pure gauge
configurations giving $n \ne 0$ are known from the physics of
instantons \cite{Wein}.

At this point, and according to our initial discussion, we are led to
associate the final local hypercharges $Y_{q}$ of Eq. (\ref{char12})
with $n=4$ as well as the current $J_{Y_{q}}$ of Eq. (\ref{equiv}) with
standard quarks.  It is important here to observe that in our model
prequarks and quarks have the same quantum numbers, except hypercharge
values. The connection is consistent with the constitutive relations of
Eq. (\ref{hyper}), the value of $n$ being independent of the
hypercharge normalization used \cite{note}.  The outcome is that initial
``bare'' prequarks transform into final ``dressed'' or normalized quarks,
which are therefore the genuine physical particles of the system.  But
prequarks are an essential part of the system because they are primary
fermions and the electroweak prequark-lepton symmetry, as inferred from
Eqs. (\ref{hyper}) and (\ref{elect}), is in the first place.  This
discrete symmetry is broken by the vacuum of the non-Abelian gauge
theory, owing to the existence of topological charge which interpolates
(in Euclidean time \cite{Huang1,Huang2}) between prequarks (at the
distant past) and quarks (at the distant future).

On the other hand, in order to make a complete scheme for the transition
from prequarks to quarks, and so effectively have the standard model, we
must add to the final hypercharge current of Eq. (\ref{equiv}) similar
relations applying to the weak and strong currents.  Consequently, we
make the following identifications in the weak sector
\begin{equation}
J^{a\mu}_{qL} = \overline{q}_{L} \gamma^{\mu} \frac{\sigma^{a}}{2}
q_{L} = \overline{\hat{q}}_{L} \gamma^{\mu} \frac{\sigma^{a}}{2}
\hat{q}_{L} = \hat{J}^{a\mu}_{qL} ,
\end{equation}
while in the color sector we identify
\begin{equation}
J^{a\mu} = \overline{q} \gamma^{\mu} \frac{\lambda^{a}}{2} q =
\overline{\hat{q}} \gamma^{\mu} \frac{\lambda^{a}}{2} \hat{q} =
\hat{J}^{a\mu} ,
\end{equation}
where $\sigma^{a}$ and $\lambda^{a}$ are the usual 2x2 Pauli and
3x3 Gell-Mann matrices, respectively.  These identifications may be
justified by noting that prequarks and quarks have the same flavor and
color quantum numbers.  Also, we include their connections in the Yukawa
sector:
\begin{eqnarray}
G^{(q)d}_{nm} \; \overline{q}_{nL} \phi \, d_{mR} &=&
\hat{G}^{(q)d}_{nm} \; \overline{\hat{q}}_{nL} \phi \, \hat{d}_{mR} ,
\nonumber \\ & & \nonumber  \\
G^{(q)u}_{nm} \; \overline{q}_{nL} \phi^{c} u_{mR} &=&
\hat{G}^{(q)u}_{nm} \; \overline{\hat{q}}_{nL} \phi^{c} \hat{u}_{mR} ,
\end{eqnarray}
where $m$=1,2,3 for the three generations, regardless of the fact that,
as in the standard model, it is not known yet how to compute Yukawa
coupling constants from first principles.

In conclusion, we have constructed step by step a Chern-Simons-fermion
model for quarks motivated by the observation that their (weak)
hypercharges are connected with leptons by precise relations, not
considered in the standard model, that may reflect a new discrete
symmetry in the electroweak sector.  The pattern shows that the quark
fractional hypercharge may be viewed as the sum of an integer ``bare''
hypercharge (being equal to the hypercharge of its lepton partner) and
a ``dressing'' or normalizing fractional hypercharge Y=4/3 which,
remarkably, is independent of the quark flavor and handness and
conserved by strong and electroweak interactions \cite{other}.
This fractional part was associated with a topological Chern-Simons
configuration of the weak gauge fields that was introduced in the
first place to cancel the gauge anomalies of the fermionic currents of
the model. The integer component of the quark hypercharge was
associated with the so-called prequark which carries the spin, flavor,
and color of the quark.  We can either regard prequarks as bare quarks
or quarks as dressed prequarks.  Original prequarks lead to an anomalous
theory while final quarks make the theory anomaly-free.  In other words,
one may conclude that the complicated collection of prequarks and
topological configurations of gauge fields behaves consistently as
single confined particles: quarks.
In some sense, we find surprising the similarity that exists between
our Chern-Simons-fermion model for quarks and the fermionic Chern-Simons
or composite-fermion theory to describe the fractional quantum
Hall effect, where electrons with local fractional electric charge are
generated from the binding of electrons of intrinsic integer charge
with a number of Chern-Simons flux quanta which screen the electron
charge \cite{composite}.  It shows at least how wide the applications
of gauge field theory combined with Chern-Simons theory can be.

\acknowledgements

We would like to thank J. Gamboa and G. Palma for useful discussions.
This work was partially supported by DICYT-USACH.


\begin{references}
\bibitem{note} For example, if the normalization $Q=T_{3}-Y$ is used
\cite{Wein}, leptons in the doublets have a fractional hypercharge 1/2,
     \linebreak
the prequark screening hypercharge becomes $Y$=-2/3, but the Pontryagin
index is still $n=4$.
\bibitem{other} Another possibility is to look for a quark-antilepton
symmetry (instead of a quark-lepton symmetry as done in this Letter)
and so have a normalizing fractional hypercharge $Y= - 2/3$, with
$Q(\hat{u})=1$ and $Q(\hat{d})=0$.  The Pontryagin index is the same
($n=4$).  We have insisted on the particle content of the standard
model.
\end{references}
\end{document}